\documentclass[preprint,aps,nofootinbib]{revtex4}
\usepackage[dvips]{graphicx}
\usepackage{amssymb}
\usepackage{amsmath}
\usepackage{amsmath, amsthm, amssymb, mathrsfs}

\begin{document}

\title{Perfect Quantum Teleportation and Superdense coding with $P_{max} = 1/2$ states}
\author{
Eylee Jung$^{1}$, Mi-Ra Hwang$^{1}$,  
DaeKil Park$^{1}$\footnote{Phone: 82-55-249-2224, \hspace{.5cm}  FAX: 82-55-244-6504, 
\hspace{.5cm} E-mail: dkpark@hep.kyungnam.ac.kr }, 
Jin-Woo Son$^{2}$, S. Tamaryan$^{3}$}

\affiliation{
$^1$ Department of Physics, Kyungnam University, Masan, 631-701, 
Korea  \\
$^2$ Department of Mathematics, Kyungnam University, Masan,
631-701, Korea         \\
$^3$ Theory Department, Yerevan Physics Institute,
Yerevan-36, 375036, Armenia}

\begin{abstract}
We conjecture that criterion for perfect quantum teleportation is that the Groverian 
entanglement of the entanglement resource is $1/\sqrt{2}$. 
In order to examine the validity of our conjecture
we analyze the quantum teleportation and superdense coding with 
$|\Phi\rangle = (1/\sqrt{2}) (|00q_1\rangle + |11q_2\rangle)$, where $|q_1\rangle$ and
$|q_2\rangle$ are arbitrary normalized single qubit states. It is shown explicitly 
that $|\Phi\rangle$ allows perfect two-party quantum teleportation and superdense
coding scenario. Next we compute the Groverian measures for 
$|\psi\rangle=\sqrt{1/2 - b^2}|100\rangle+b |010\rangle+a|001\rangle
+\sqrt{1/2-a^2}|111\rangle$
and $|\tilde{\psi}\rangle=a|000\rangle+b|010\rangle+\sqrt{1/2 - (a^2+b^2)}|100\rangle 
+ (1/\sqrt{2}) |111\rangle$, 
which also allow the perfect quantum teleportation. It is shown that both states have 
$1/\sqrt{2}$ Groverian entanglement measure, which strongly supports that our conjecture
is valid.  
\end{abstract}


\maketitle

Quantum teleportation\cite{bennett93} is a physical process, where an unknown state
can be transmitted from one remote place to another by making use of the entanglement
resource
and classical communication. About one and half decades ago Bennett et al\cite{bennett93} 
have found such process. They used the two-qubit Einstein-Podolsky-Rosen (EPR) state as
an entanglement resource, which is assumed to be initially shared between the sender, 
called 
Alice and the receiver, called Bob. Quantum teleportation is generalized to the case
where the noisy channels make a quantum channel to be mixed state\cite{oh02}. In this 
case the quantum teleportation generally becomes imperfact due to the effect of the 
noisy channels. 
If we do not have EPR state or its local-unitary(LU) equivalents, then Alice cannot 
teleport a single qubit to Bob with unit fidelity and unit probability. Increasing the
fidelity as much as possible, one can achieve an unit fidelity with a probability less
than a unit, which is called probabilistic quantum 
teleportation\cite{agra02-1,gordon06-1}. 

The higher-qubit entangled states also can be used as an entanglement resource
of the quantum 
teleportation. In $n$-qubit
system with $n \geq 3$ there seems to be no unique way to define the maximally 
entangled states. For $n=3$, for example, it is well-known that there are two 
types of entangled states called Greenberger-Horne-Zeilinger(GHZ) state\cite{green89}
\begin{equation}
\label{ghz1}
|GHZ \rangle = \frac{1}{\sqrt{2}} \left( |000 \rangle + |111 \rangle \right)
\end{equation}
and W state\cite{dur00-1}
\begin{equation}
\label{W1}
|W \rangle = \frac{1}{\sqrt{3}} \left( |001\rangle + |010\rangle + |100\rangle \right).
\end{equation}
These two types are not connected to each other via stochastic local operations and 
classical communication(SLOCC)\cite{dur00-1}. It was also found that four qubits can
be entangled in nine different ways\cite{verstra02}. 

We generally use the entanglement measures to quantify the entanglement of multi-qubit
state $|\psi\rangle$. 
One of the well-known measure constructed by an operational 
method\footnote{
In operational method the entanglement measures are constructed by making use of 
the real physical tasks such as quantum algorithms. In fact, the Groverian 
entanglement measure was constructed from Grover's search algorithm.}
is a 
Groverian measure\cite{biham01-1} defined $G(\psi) \equiv \sqrt{1 - P_{max}}$, where 
\begin{equation}
\label{pmax1}
P_{max} = \max_{|e_1\rangle \cdots |e_n\rangle}
|\langle e_1| \otimes \cdots \otimes \langle e_n| \psi \rangle|^2.
\end{equation}
Physically, $P_{max}$ corresponds to the maximal probability of success in 
Grover's search algorithm\cite{grover97} when $|\psi\rangle$ is $n$-qubit initial 
state. Eq.(\ref{pmax1}) can be re-written in terms of density matrix 
$\rho = |\psi\rangle \langle \psi |$ in the form
\begin{equation}
\label{pmax2}
P_{max} = \max_{R^1 \cdots R^n} \mbox{Tr} 
\left[\rho R^1 \otimes \cdots \otimes R^n \right]
\end{equation}
where $R^i \equiv |q_i \rangle \langle q_i |$.

The quantum teleportation with $3$-qubit GHZ state was discussed in Ref.\cite{karl98}.
When one sender (Alice) would like to send one-qubit state
\begin{equation}
\label{comm1}
|\tilde{\psi}\rangle_1 = \alpha |0\rangle + \beta |1\rangle
\hspace{2.0cm} (|\alpha|^2 + |\beta|^2 = 1)
\end{equation}
to one receiever (Bob), the perfect quantum teleportation with $|GHZ\rangle_{234}$ can
be easily shown as following. First, we assume that Alice has particles $2$ and $3$, 
and Bob has particle $4$. Next, we note that 
$|\tilde{\psi}\rangle_1 \otimes |GHZ\rangle_{234}$ reduces to
\begin{eqnarray}
\label{3qghz-1}
& &|\tilde{\psi}\rangle_1 \otimes |GHZ\rangle_{234} 
= \bigg[ \sqrt{P_1^+} |\phi_1^+ \rangle_{123} \otimes \openone + 
         \sqrt{P_1^-} |\phi_1^- \rangle_{123} \otimes Z    \\   \nonumber
& & \hspace{2.0cm} +
         \sqrt{P_2^+} |\phi_2^+ \rangle_{123} \otimes X + 
         \sqrt{P_2^-} |\phi_2^- \rangle_{123} \otimes Z X \bigg]
      \left(\alpha |0\rangle_4 + \beta |1\rangle_4 \right)
\end{eqnarray}
where $P_1^+ = P_1^- = P_2^+ = P_2^- = 1/4$, ($X$, $Y$, $Z$) Pauli matrices, and
\begin{eqnarray}
\label{3qghz-2}
& & |\phi_1^{\pm}\rangle = 
                      \frac{1}{\sqrt{2}} \left( |000\rangle \pm |111\rangle
                                                          \right)  \\  \nonumber
& & |\phi_2^{\pm}\rangle = 
                      \frac{1}{\sqrt{2}} \left( |100\rangle \pm |011\rangle
                                                          \right).
\end{eqnarray}
Since $|\phi_1^{\pm}\rangle$ and $|\phi_2^{\pm}\rangle$ are orthogonal to each others,
Alice can distinguish them via von Neumann type measurement. Of course, the postulates
of quantum mechanics tells that the probabilities for outcomes are 
$P_1^{\pm}$ and $P_2^{\pm}$, respectively. After Alice conveys her measurement 
results to Bob via classical channel, Bob can construct $|\tilde{\psi}\rangle$ by 
applying an appropriate unitary transformation to his own qubit. This is a whole
story of quantum teleportation between two parties. 
                                                        
Since the quantum teleportation between two parties can be done perfectly with the
two-qubit EPR channel, actually the above-mentioned teleportation is not new 
scheme. However, the three-qubit GHZ state can be used to three-party (Alice, Bob,
Cliff) teleportation. Although the well-known no-cloning 
theorem\cite{wootters82,barnum96-1} does not allow for Alice to teleport 
$|\tilde{\psi} \rangle$ to both Bob and Cliff, one can use the $3$-qubit GHZ state as a 
quantum copier (cloning device)\cite{buzek97-1,gisin97,buzek96-1} with fidelity less
than one\cite{karl98}. The quantum teleportation with four qubit GHZ state and its
role as a cloning machine was discussed in Ref.\cite{pati00-1}. 

Recently, furthermore,
the slightly-modified W state 
\begin{equation}
\label{W-1}
|W_1 \rangle_{234} = \frac{1}{2} \left( |100 \rangle_{234} + |010 \rangle_{234} + 
               \sqrt{2} |001 \rangle_{234} \right)
\end{equation}
is used for perfect two-party quantum teleportation\cite{agra06-1}.
This can be shown as following. First let us assume that Alice has particles 
$2$ and $3$, and Bob has particle $4$. Then after some calculation it is easy to 
show
\begin{eqnarray}
\label{W-teleport}
& &|\tilde{\psi}\rangle_1 \otimes |W_1\rangle_{234} 
= \sqrt{\frac{1}{4}} \bigg[ |\eta_1^+\rangle_{123} \otimes \openone + 
                            |\eta_1^-\rangle_{123} \otimes Z
                                                              \\   \nonumber
& &\hspace{2.0cm} +        |\xi_1^+\rangle_{123} \otimes X + 
                           |\xi_1^-\rangle_{123} \otimes ZX \bigg]
                (\alpha |0\rangle_4 + \beta |1\rangle_4)
\end{eqnarray}
where
\begin{eqnarray}
\label{W-2}
& &|\eta_1^{\pm}\rangle = \frac{1}{2} \left( |010\rangle + |001\rangle \pm 
                          \sqrt{2} |100\rangle \right)
                                                           \\  \nonumber
& &|\xi_1^{\pm}\rangle = \frac{1}{2} \left( |110\rangle + |101\rangle \pm 
                          \sqrt{2} |000\rangle \right).
\end{eqnarray}
Since $|\eta_1^{\pm}\rangle$ and $|\xi_1^{\pm}\rangle$ are orthogonal to each others,
the usual quantum teleportation process allows Bob to have $|\tilde{\psi}\rangle$ via
an appropriate unitary transformation. The difference of this process 
from teleportation with 
$|GHZ\rangle$ is that in this case Alice should initially 
choose particles $2$ and $3$ for 
perfect teleportation. If Alice has different particles, one can show that the perfect
teleportation is impossible with state $|W_1\rangle$. Since, however, initially 
Alice and Bob can choose particles freely, we can use $|W_1\rangle$ for perfect
two-party quantum teleportation.

The perfect teleportation with $|W_1\rangle$ and $|GHZ\rangle$ naturally arises a 
question: what is a criterion for the perfect two-party quantum teleportation? In 
other words what common property of $|W_1\rangle$ and $|GHZ\rangle$ allows perfect
teleportation? As will be shown below, $|W_1\rangle$ and $|GHZ\rangle$ have 
same $P_{max} = 1/2$. Thus this fact might be criterion for the perfect teleportation.
The purpose of this paper is to explore this issue in detail.

Recently, it was shown\cite{jung07-1} that $P_{max}$ for $n$-qubit state can be 
computed if one knows one of the $(n-1)$-qubit reduced states using a formula
\begin{equation}
\label{theorem1}
P_{max} = \max_{R^1 \cdots R^n} \mbox{Tr} \left[\rho R^1 \otimes \cdots \otimes R^n
                                                                \right]
= \max_{R^1 \cdots R^{n-1}} \mbox{Tr} \left[\rho R^1 \otimes \cdots \otimes R^{n-1}
                                        \otimes \openone \right].
\end{equation}
Eq.(\ref{theorem1}) leads several important conclusions\cite{jung07-1}. Furthermore,
Eq.(\ref{theorem1}) provides a good tool for the analytic calculation of $P_{max}$.
In Ref.\cite{tamaryan07-1} $P_{max}$ for various $3$-qubit states was analytically
computed using Eq.(\ref{theorem1}). For the generalized W state, for example,
\begin{equation}
\label{gw-1}
|GW \rangle = a |001\rangle + b |010\rangle + c |100\rangle
\hspace{2.0cm} (a^2 + b^2 + c^2 = 1)
\end{equation}
$P_{max}$ can be expressed as following:
\begin{eqnarray}
\label{pmaxgw}
P_{max} = \left\{    \begin{array}{ll}
         \max (a^2, b^2, c^2) = \alpha^2  &  \hspace{1.0cm} \mbox{when} \hspace{.2cm} 
                                              \alpha^2 \geq \beta^2 + \gamma^2   \\
                     4 R^2             &  \hspace{1.0cm} \mbox{when} \hspace{.2cm}      
                                              \alpha^2 \leq \beta^2 + \gamma^2  
                      \end{array}            \right.
\end{eqnarray}
where $\alpha^2 = \max (a^2, b^2, c^2)$ and, $\beta^2$ and $\gamma^2$ are the remaining
ones. In Eq.(\ref{pmaxgw}) $R$ is a circumradius of the triangle $a$, $b$, $c$. 
From Eq.(\ref{pmaxgw}) it is easy to show that if $a$, $b$, $c$ form an equilateral
triangle, $P_{max} = 4 / 9$, which is consistent with the results 
of Ref.\cite{shim04-1}.
Furthermore, Eq.(\ref{pmaxgw}) implies that if the parameters $a$, $b$, $c$ form a 
right triangle, we call the corresponding $|GW\rangle$ `singular states'\footnote{The
suitability of the terminology `singular states' can be seen easily if one changes 
Eq.(\ref{gw-1}) into the one-parameter dependent states by letting $b = \kappa a$ and 
$c = \kappa^2 a$. Then Eq.(\ref{pmaxgw}) allows oneself to derive the analytic 
expressions for the $\kappa$-dependence of $P_{max}$. Using these expressions,  
one can show that 
$P_{max}$ at right triangle $a$, $b$, $c$ is continuous but its derivative 
$d P_{max} / d \kappa$ is discontinuous. For general state (\ref{gw-1}) the 
normalization condition $a^2+b^2+c^2=1$ defines a sphere and the condition for right 
triangle $\alpha^2 = \beta^2 + \gamma^2$ defines a cone. The intersection of the cone
with the sphere is generally a circle on the sphere. Inside the circle 
$P_{max} \leq 1/2$ and outside $P_{max} \geq 1/2$. On the circle $P_{max} = 1/2$ but
its gradient is discontinuous. Thus it is reasonable to use the terminology 
`singular states' for the states with $\alpha^2 = \beta^2 + \gamma^2$.} 
and their $P_{max}$ becomes $1/2$. Since 
$|W_1\rangle$ in Eq.(\ref{W-1}) is one of singular states, 
its $P_{max}$ is $1/2$ as we commented
before. Since it is well-known that the $n$-qubit GHZ states has $P_{max}=1/2$ 
regardless of $n$\cite{shim04-1}, this remarkable fact makes us conjecture that
$P_{max}=1/2$ is a necessary (or sufficient) condition for the perfect two-party
teleportation.

To examine the validity of our conjecture we choose the state
\begin{equation}
\label{ours1}
|\Phi\rangle_{234} = \frac{1}{\sqrt{2}} \left( |00q_1\rangle + |11q_2\rangle \right)
\end{equation}
where $|q_1\rangle$ and $|q_2\rangle$ are arbitrary normalized one-qubit states. If 
$|q_1\rangle = |0\rangle$ and $|q_2\rangle = |1\rangle$, $|\Phi\rangle$, of course,
becomes usual GHZ state. As shown in Ref.\cite{tamaryan07-1}, $P_{max}$ of 
$|\Phi\rangle$ is also $1/2$ regardless of $|q_1\rangle$ and $|q_2\rangle$. 
Thus this state is appropriate to check the 
validity of our conjecture. We will show that like GHZ and W states $|\Phi\rangle$ also
allows the perfect quantum teleportation and superdense coding scenario. Next, we will
compute $P_{max}$ of more general three-qubit states, 
which also allow the perfect teleportation.
As we conjecture, it is shown that these general states also have $P_{max} = 1/2$.

In order to discuss the two-party quantum teleportation with $|\Phi\rangle$ we assume 
first that Alice has particles $3$ and $4$, and Bob has particle $2$. In this 
situation it is convenient to define
\begin{eqnarray}
\label{3qours-1}
& &|\psi_1^{\pm} \rangle = \frac{1}{\sqrt{2}} \left[ |00q_1\rangle \pm |11q_2\rangle
                                                           \right]   \\   \nonumber
& &|\psi_2^{\pm} \rangle = \frac{1}{\sqrt{2}} \left[ |10q_1\rangle \pm |01q_2\rangle
                                                           \right].
\end{eqnarray}
Then one can show that $|\psi_1^{\pm}\rangle$ and $|\psi_2^{\pm}\rangle$ are 
orthogonal to each others regardless of $|q_1\rangle$ and $|q_2\rangle$. After some
calculation one can show straightforwardly that 
$|\tilde{\psi}\rangle_1 \otimes |\Phi\rangle_{234}$ reduces to
\begin{eqnarray}
\label{3qours-2}
& &|\tilde{\psi}\rangle_1 \otimes |\Phi\rangle_{234} = 
\sqrt{\frac{1}{4}} \bigg[ |\psi_1^+\rangle_{134} \otimes \openone + 
|\psi_1^-\rangle_{134} \otimes Z           \\   \nonumber 
& & \hspace{2.0cm}
+ |\psi_2^+\rangle_{134} \otimes X + 
|\psi_2^-\rangle_{134} \otimes ZX \bigg] (\alpha |0\rangle_2 + \beta |1\rangle_2).
\end{eqnarray}
Thus Alice can send $|\tilde{\psi}\rangle$ to Bob via usual quantum teleportation 
process: she distinguishes $|\psi_1^{\pm}\rangle$ and $|\psi_2^{\pm}\rangle$ via
von Neumann type measurement and conveys her measurement outcomes to Bob via classical 
channel. If Bob has particle $3$ and Alice has particles $2$ and $4$, one can show 
similarly that a perfect quantum teleportation with $|\Phi\rangle$ is also possible. 

Finally, let us consider the situation that Bob has particle $4$ and Alice has particles
$2$ and $3$. Even in this case one can show that perfect quantum teleportation is 
possible if $|q_1\rangle$ is orthogonal to $|q_2\rangle$, {\it i.e.} 
$\langle q_1 | q_2 \rangle = 0$. If $\langle q_1 | q_2 \rangle = 0$, there should 
exist an unitary operator $u$ such that $|q_1\rangle = u |0\rangle$ and 
$|q_2\rangle = u |1\rangle$ because unitary operator preserves the inner product. Then
$|\Phi\rangle$ is obtained from $|GHZ\rangle$ via local-unitary transformation
as following
\begin{equation}
\label{localu-1}
|\Phi\rangle = \left(\openone \otimes \openone \otimes u\right) |GHZ\rangle.
\end{equation}
Then Eq.(\ref{3qghz-1}) implies 
\begin{eqnarray}
\label{3qours-3}
& &|\tilde{\psi}\rangle_1 \otimes |\Phi\rangle_{234} 
= \sqrt{\frac{1}{4}} \bigg[ |\phi_1^+ \rangle_{123} \otimes  u \openone +
         |\phi_1^- \rangle_{123}  \otimes u Z    \\   \nonumber
& & \hspace{2.0cm} +
         |\phi_2^+ \rangle_{123} \otimes u X +
         |\phi_2^- \rangle_{123} \otimes u Z X \bigg]
      \left(\alpha |0\rangle_4 + \beta |1\rangle_4 \right).
\end{eqnarray}
Therefore, if Alice has outcome $|\phi_1^+ \rangle$ via her measurement, Bob can get
$|\tilde{\psi}\rangle$ by operating $u^{-1} = u^{\dagger}$ to his qubit. If she has
$|\phi_1^- \rangle_{123}$, $|\phi_2^+ \rangle_{123}$ and 
$|\phi_2^- \rangle_{123}$ respectively,
Bob should operate $Z u^{-1}$, $X u^{-1}$ and $X Z u^{-1}$ for each case to get 
$|\tilde{\psi}\rangle$. Thus perfect quantum teleportation is possible.
Although perfect two-party quantum teleportation is impossible provided that Bob has
initially particle $4$ and $\langle q_1 | q_2 \rangle \neq 0$, we can use 
$|\Phi\rangle$ for perfect teleportation because initially Alice and Bob can choose 
their particles freely. This is exactly same situation with teleportation with 
$|W_1\rangle$. In conclusion we can use $|\Phi\rangle$ for the perfect two-party 
quantum teleportation. This strongly supports our conjecture that {\it the criterion 
for the perfect quantum teleportation is $P_{max} = 1/2$}.

Next we would like to discuss the superdense coding\cite{bennett92} 
with $|\Phi\rangle$. In  
order for the superdense coding scenario to work Alice should be able to send 
two classical bits to Bob by sending one qubit. Now we assume that Alice has particle 
$2$ and Bob has particles $3$ and $4$ in $|\Phi\rangle_{234}$. If Alice applies 
($\openone$, $Z$, $X$, $-iY$) to her qubit, $|\Phi\rangle$ changes into
\begin{eqnarray}
\label{dense1}
& &\left(\openone \otimes \openone \otimes \openone \right) |\Phi\rangle =
                   |\psi_1^+ \rangle                    \\   \nonumber
& &\left(Z \otimes \openone \otimes \openone \right) |\Phi\rangle =
                   |\psi_1^- \rangle                    \\   \nonumber
& &\left(X \otimes \openone \otimes \openone \right) |\Phi\rangle =
                   |\psi_2^+ \rangle                    \\   \nonumber
& &\left(-iY \otimes \openone \otimes \openone \right) |\Phi\rangle =
                   |\psi_2^- \rangle                  
\end{eqnarray}
respectively. Since $|\psi_1^{\pm}\rangle$ and $|\psi_2^{\pm}\rangle$ are orthogonal
to each other, Bob can distinguish them via von Neumann type measurement if Alice send
her one qubit to him. This completes the superdense coding scenario with 
$|\Phi\rangle$. Similarly, one can show that we can complete the superdense coding
scenario if Alice has particle $3$ and Bob has particles $2$ and $4$. Although perfect
superdense coding scenario is impossible provided that Alice has particle $4$ and 
$\langle q_1 | q_2 \rangle \neq 0$, we can use $|\Phi\rangle$ for prefect superdense
coding because initially Alice and Bob can share particles freely at their convenience.

Recently, two three-qubit states were found, which allow the perfect 
quantum teleportation\cite{xin07}. We would like to show that both states also
have $P_{max} = 1/2$ as we conjecture. This strongly supports the validity of our 
conjecture again. First state is 
\begin{equation}
\label{first1}
|\psi\rangle = \sqrt{\frac{1}{2} - b^2} |100\rangle + b |010\rangle + a |001\rangle
              + \sqrt{\frac{1}{2} - a^2} |111\rangle.
\hspace{1.0cm}
\left(0 \leq a, b \leq \frac{1}{\sqrt{2}} \right)
\end{equation}
If $a = 1/\sqrt{2}$ and $b=1/2$, $|\psi\rangle$ reduces to $|W_1\rangle$ 
defined in Eq.(\ref{W-1}).
Let $\{\alpha, \beta, \gamma, \delta \}$ be set of 
$\{a, b, \sqrt{1/2 - a^2}, \sqrt{1/2 - b^2} \}$ with decreasing order. Then one can show 
easily $\alpha^2 \leq \beta^2 + \gamma^2 + \delta^2 + 2 \beta \gamma \delta / \alpha$
regardless of $a$ and $b$. As shown in Ref.\cite{tama08-1}, then, $P_{max}$ for 
$|\psi\rangle$ equals to $4 R^2$, where $R$ is a circumradius of convex quadrangle:
\begin{equation}
\label{first2}
R^2 = \frac{(a_1 a_2 + a_3 a_4) (a_1 a_3 + a_2 a_4) (a_1 a_4 + a_2 a_3)}
           {4 \omega^2 - r_3^2}
\end{equation}
where $\omega = a_1 a_2 + a_3 a_4$, $r_3 = a_1^2 + a_2^2 - a_3^2 - a_4^2$, and 
the constants $a_i$'s are the coefficients of the quantum channel (\ref{first1}).
If we put $a_1 = \sqrt{1/2 - b^2}$, $a_2 = b$, $a_3=a$ and $a_4 = \sqrt{1/2 - a^2}$,
one can show easily that $P_{max}$ for $|\psi\rangle$ in Eq.(\ref{first1}) is $1/2$.

Second state which allows a perfect quantum teleportation is 
\begin{equation}
\label{second1}
|\tilde{\psi}\rangle = a |000\rangle + b |010\rangle + \sqrt{\frac{1}{2} - (a^2 + b^2)}
|100\rangle + \frac{1}{\sqrt{2}} |111\rangle.
\hspace{1.0cm}
\left((0 \leq a^2 + b^2 \leq \frac{1}{2} \right)
\end{equation}
If $a=1/\sqrt{2}$ and $b=0$, $|\tilde{\psi}\rangle$ exactly coincides with GHZ state.
As shown in Ref.\cite{jung07-1,tamaryan07-1}, $P_{max}$ for $|\tilde{\psi}\rangle$ can
be writtten as 
\begin{equation}
\label{second2}
P_{max} = \max_{|\vec{s}_2| = |\vec{s}_3| = 1}
\frac{1}{4} \left[1 + \vec{s}_2 \cdot \vec{r}_2 + \vec{s}_3 \cdot \vec{r}_3 + 
                  s_{2i} s_{3j} g_{ij} \right]
\end{equation}
where
\begin{eqnarray}
\label{second3}
& &\vec{r}_2 = \mbox{Tr} [\rho^B \vec{\sigma} ] = (2 a b, 0, -2b^2)
                                                          \\   \nonumber
& &\vec{r}_3 = \mbox{Tr} [\rho^C \vec{\sigma} ] = (0, 0, 0)
                                                          \\   \nonumber
& &g_{ij} = \mbox{Tr} [\rho^{BC} \sigma_i \otimes \sigma_j] = 
\left(               \begin{array}{ccc}
         \sqrt{1 - 2(a^2 + b^2)}   &   0   &   2 a b         \\
         0   &   -\sqrt{1 - 2(a^2 + b^2)}   &   0            \\
         0   &   0   &   1 - 2 b^2
                     \end{array}                                \right).
\end{eqnarray}
In Eq.(\ref{second3}) $\rho^{BC}$, $\rho^B$ and $\rho^C$ are the corresponding 
partial traces of $\rho^{ABC} \equiv |\tilde{\psi}\rangle \langle \tilde{\psi}|$ and 
$\sigma_i$'s are usual Pauli matrix. Due to maximization in Eq.(\ref{second2}), 
$\vec{s}_2$ and $\vec{s}_3$ satisfy the Lagrange multiplier equations
\begin{eqnarray}
\label{second4}
& &\vec{r}_2 + g \vec{s}_3 = \Lambda_1 \vec{s}_2
                                                 \\   \nonumber
& &\vec{r}_3 + g^T \vec{s}_2 = \Lambda_2 \vec{s}_3
\end{eqnarray}
with $\Lambda_1, \Lambda_2 > 0$. Let $\vec{s}_2 = (s_{2x}, s_{2y}, s_{2z})$ and 
$\vec{s}_3 = (s_{3x}, s_{3y}, s_{3z})$. Then Lagrange multiplier equations in general
reduce to six-degree algebraic equations and it is usually impossible to derive the 
solutions analytically. For $|\tilde{\psi}\rangle$, however, Eq.(\ref{second4}) reduce
to simple cubic equations due to dramatic cancellation between left and right sides.
Due to this cancellation we can derive analytic solutions which are 
$s_{2y} = s_{3y} = 0$ and 
\begin{equation}
\label{second5}
s_{3z} = \frac{(a^2 - b^2) + 2 b^2 (a^2 + b^2)}{(a^2 + b^2) (1 - 2 b^2)}
\hspace{1.0cm}   \mbox{or} \hspace{1.0cm}
\pm b \sqrt{\frac{1 - 2 (a^2 + b^2)}{(a^2 - b^2) (1 - 2 b^2)}}.
\end{equation}
The first solution of Eq.(\ref{second5}) yields the remaining solutions
\begin{equation}
\label{second6}
s_{3x} = \frac{2 a b \sqrt{1 - 2(a^2 + b^2)}}{(a^2 + b^2) (1 - 2 b^2)}
\hspace{1.0cm}
s_{2x} = \frac{2 a b}{a^2 + b^2}
\hspace{1.0cm}
s_{2z} = \frac{a^2 - b^2}{a^2 + b^2}
\end{equation}
and positive Lagrange multiplier constants $\Lambda_1 = 1$ and $\Lambda_2 = 1 - 2b^2$.
Then Eq.(\ref{second2}) gives $P_{max} = 1/2$. The second solution of Eq.(\ref{second5})
also yields the different remaining solutions, but the corresponding $P_{max}$ is 
$(1/4) (1 + \sqrt{1 - 2 a^2})$, which is smaller than $1/2$. Since we should take 
maximization in Eq.(\ref{second2}), $P_{max}$ for $|\tilde{\psi}\rangle$ should be $1/2$.

We have shown that $|\Phi\rangle$, whose $P_{max}$ is $1/2$, allows 
the perfect two-party quantum teleportation. Also we have shown that $|\psi\rangle$ and
$|\tilde{\psi}\rangle$ in Eq.(\ref{first1}) and Eq.(\ref{second1}) have 
$P_{max} = 1/2$. The usual GHZ and W states are special limits of $|\psi\rangle$ and
$|\tilde{\psi}\rangle$. This means that our conjecture ``{\it the criterion for the 
perfect two-party quantum teleportation is $P_{max} = 1/2$}'' is widely applicable.
Since we cannot find any counter-example, we feel that this criterion is a 
necessary and sufficient condition. In other words ``{\it the perfect two-party
quantum teleportation is possible if and only if the Groverian measure for the 
entanglement resource is $1/\sqrt{2}$}''. But more rigorous proof is needed for 
this statement.

{\bf Acknowledgement}: 
This work was supported by the Kyungnam University
Research Fund, 2007.

\end{document}